# Calibration of free-space and fiber-coupled single-photon detectors

Thomas Gerrits[1], Alan Migdall[2], Joshua C. Bienfang[2], John Lehman[1], Sae Woo Nam[1], Jolene Splett[1], Igor Vayshenker[1], Jack Wang[1]

[1] National Institute of Standards and Technology, Boulder, CO, 80305, USA
[2] Joint Quantum Institute, University of Maryland and National Institute of Standards and Technology, Gaithersburg, MD, 20899, USA

### Abstract

We measure the detection efficiency of single-photon detectors at wavelengths near 851 nm and 1533.6 nm. We investigate the spatial uniformity of one free-space-coupled single-photon avalanche diode and present a comparison between fusion-spliced and connectorized fiber-coupled single-photon detectors. We find that our expanded relative uncertainty for a single measurement of the detection efficiency is as low as 0.70 % for fiber-coupled measurements at 1533.6 nm and as high as 1.78 % for our free-space characterization at 851.7 nm. The detection-efficiency determination includes corrections for afterpulsing, dark count, and count-rate effects of the single-photon detector with the detection efficiency interpolated to operation at a specified detected count rate.

## 1. Introduction

Detection of light is an enabling technology for many applications and current detection capabilities are impressive, covering a dynamic range of ≈20 orders of magnitude, from just a few femtowatts to 100's of kilowatts of optical power. Kilowatts of power can now accurately be measured by use of the photon momentum of optical beams, a convenient method that 'weighs' the optical power on a scale, after which the optical mode can still be used for an experiment or application [1]. On the other end of the optical power scale are applications driving advances in single-photon-counting technologies such as: phase discrimination, Bell tests, exotic quantum states of light, low-light imaging and ranging, etc. [2-8]. Accurate knowledge of a single-photon detector's efficiency is a prerequisite for many of these applications, particularly those that rely on quantum effects. Also, single-photon counting offers the unique capability of measuring optical power by counting photons, a regime distinct from analog measurements and one that offers the potential for inherently higher accuracy. To date, no such photon-counting-based standard exists. However, the international system of units (SI) will soon be recast based on fundamental constants and laws of nature [9, 10]. Part of the new quantum SI could be a source- or detector-based single-photon standard. For this reason, many national metrology institutes around the world are pursuing the establishment of single-photon-based traceable or absolute calibrations of single-photon detectors and sources.

Low-uncertainty measurements of the detection efficiency (DE) of a single-photon detector (SPD) are challenging. Detection efficiency is defined as the probability of detecting a photon incident on the detector, as distinct from the other quantities such as quantum efficiency that relate to just a portion of the detection process. One common method of calibrating an SPD is by use of an attenuated laser source [11, 12]. This measurement requires accurate knowledge of the laser power at microwatt levels or lower, achieved via a calibrated optical power meter traceable to a primary standard. Attenuation of the laser power to the single-photon regime is achieved by calibrating attenuator(s) over multiple orders of magnitude. Müller *et al*. have demonstrated a different method for SPD calibration by use of a synchrotron light source [13, 14]. The synchrotron output flux is linear with ring current, thus by control and measurement of the ring current the synchrotron's output can be tuned over many orders of magnitude, extending even to single-photon levels, without the need of attenuator calibration. Yet another method uses a correlated photon source such as those based on spontaneous parametric downconversion [15-18]. That method has the additional feature that it is inherently absolute, albeit for the efficiency of the entire source-to-detector system. Thus, to determine the detector's portion of the overall efficiency, additional measurements are required, such as the losses of the optical path from the source to the detector of interest.

Here, we report on traceable calibrations of SPD detection efficiencies. Our method employs optical power meters calibrated at high power levels ($\approx\mu$W) that can maintain high accuracy at low light levels ($\approx$pW) [19, 20]. We use a calibrated beam splitter and a monitor power meter in combination with an optical fiber attenuator to extend our measurement scale to levels compatible with single-photon detectors. The method allows us to accurately control the photon flux at the detector under test (DUT) with an uncertainty dominated by the calibration of our optical power meters.

We measured the detection efficiencies of four detectors: one free-space silicon single-photon avalanche diode (SPAD), two optical fiber-coupled Si-SPADs, and one superconducting nanowire single-photon detector (SNSPD), all at a wavelength of $\approx$851 nm. We present our methods and associated uncertainties at a specified single-photon count rate and wavelength. We also report our measurement results for one fiber-coupled SNSPD at a wavelength of 1533.6 nm. In addition, to quantify the effect of fiber-to-fiber connections on the DE measurement, we made measurements of the SNSPD DE employing both commercial ferule connector/physical contact (FC/PC) fiber connectors and fusion-spliced connections. The methods presented here represent a straightforward and accessible effort to accurately characterize DE using standard technologies. We highlight some of the challenges unique to characterizing free-space and fiber-coupled single photon detectors, and achieve overall uncertainties based on absolute methods.

## 2. Experimental methods

All our measurements and calibrations are made using the experimental scheme shown in Figure 1. Laser light through a variable fiber attenuator (VFA$_{input}$) is sent to the splitter/attenuator unit where the input is monitored and the output-to-monitor ratio ($R_{\text{out/mon}}$) of $\approx 10^{-5}$ is measured using our calibrated power meter (PM) and monitor power meter (PM$_{mon}$). Key to the measurements are the transmittance of the splitter/attenuator unit and the output-to-monitor ratio of the splitter/attenuator unit. Both are determined from the fiber beam splitter (FBS) splitting ratio and the attenuation of VFA, as measured using the calibrated power meter and the monitor power meter. In addition, this method relies on the stability of the splitter/attenuator unit's output-to-monitor ratio, the polarization and wavelength of the light versus time, and the independence of the output-to-monitor ratio with input optical power. We verify each of these either during the measurement or by prior characterization of the setup components.

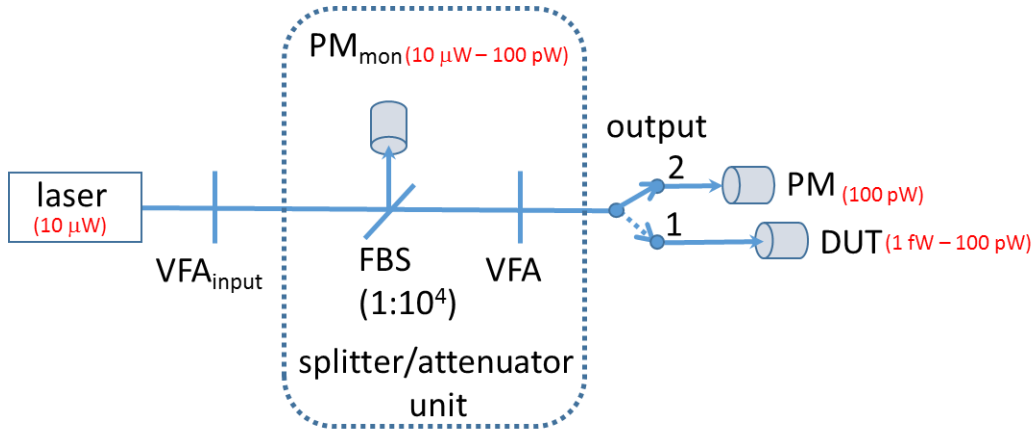

Figure 1: A schematic of the setup used throughout this study. A fiber-coupled laser is coupled to a variable fiber attenuator (VFA$_{input}$) followed by a beam splitter/attenuator unit consisting of a monitor power meter (PM$_{mon}$), a fiber beam splitter (FBS) with a 1:10$^4$ split ratio, and another variable fiber attenuator (VFA). Switching the output controls whether the light goes to the calibrated power meter (PM) or the detector under test (DUT).

The output-to-monitor ratio is measured at high light levels (with VFA$_{input}$ set for low attenuation) using PM and PM$_{mon}$. Then the input power is reduced by VFA$_{input}$ to put the output power in the desired range for the DUT. Thus, knowledge of the output-to-monitor ratio and the measured power at PM$_{mon}$ allows absolute determination of the optical power or single-photon flux at the DUT. VFA$_{input}$ only serves as a power dial at the DUT and does not have to be calibrated. We also note that the dynamic range of our calibrated optical power meters is 8 orders of magnitude, allowing high accuracy measurements of that ratio.

For example, an optical power of ≈10 µW is coupled into the optical fiber before VFA$_{input}$. After VFA$_{input}$ the light is launched into the FBS with splitting ratio 1:10$^4$. The high-power output port is monitored by PM$_{mon}$ [20]. In this example we set VFA to -10 dB, increasing the output-to-monitor ratio to 1:10$^5$. The light is then directed to PM [19]. If the input power from the laser is ≈10 µW, the optical power at the PM will be ≈100 pW. Several measurements of the output-to-monitor ratio are performed before each DUT measurement. With VFA kept constant, we adjust VFA$_{input}$ to a value compatible with the dynamic range of the DUT. In this example, we adjust VFA$_{input}$ to -40 dB. This results in an optical power of ≈10 fW at the DUT, or ≈43000 photons per second and 77000 photons per second incident at the DUT for wavelengths 851 nm and 1533.6 nm, respectively. When switching wavelength, we use single-mode fiber for the wavelength of interest. The VFAs are broad-band and cover the region from 750 nm to 1700 nm. For wavelengths below 1200 nm, the VFAs are multimode. This could impact the attenuation setting repeatability at lower wavelengths. However, we are not relying on the repeatability of our VFAs, since we are measuring the output-to-monitor ratio each time before a measurement run. We measured the output-to-monitor ratio for a range of incoming optical powers and the results are presented in the next section. In addition, we monitor the temperature, laser wavelength, and polarization, as they may also affect the output-to-monitor ratio.

Further, switching the PM with the DUT (substitution method) is done in three ways. For our free-space measurements we mount the PM along with the DUT on an automated *xyz*-translation stage and move each of the detectors into the beam path. For our fiber-based measurements we disconnect the PM and connect the DUT either by use of a FC/PC fiber connector union (adapter) or by breaking and re-splicing the optical fibers.

## 2.1 Output-to-monitor ratio stability for 851 nm and 1533 nm

To verify that the output-to-monitor ratio is independent of the optical input power, we measured it for a range of input powers by adjusting $VFA_{input}$. Figure 2 shows the variation of the splitting ratio versus attenuator setting for two wavelengths. From the data, we calculated a relative expanded uncertainty ($k = 2$) of 0.4 % and 0.1 % for 851.8 nm and 1533.6 nm, respectively. The output-to-monitor ratio is within the combined uncertainty of the nonlinearity correction and the measurement uncertainty at both wavelengths. The jump in ratio seen in Figure 2(b) at $VFA_{input} \approx 40$ dB is due to a power meter range change. The error bars are dominated by the calibration uncertainties.

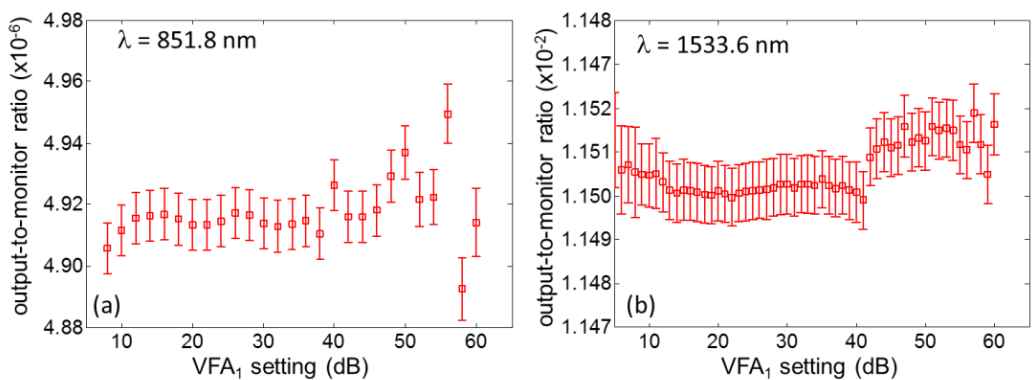

Figure 2: Measured output-to-monitor ratio versus $VFA_{input}$ setting. (a) for the 851 nm setup. (b) for the 1533.6 nm setup. The error bars in both figures are dominated by the nonlinearity correction uncertainties ($k=2$).

## 2.2 Si-Trap detector as calibrated power meter

We used a Ti:sapphire oscillator with $\approx 5$ nm bandwidth as one of the photon sources and a Si-Trap detector (SiTrap) as the PM for all free-space measurements [21]. In addition, the reflectivity of the SiTrap is less than 1 %, and its spatial response nonuniformity is extremely small, ($<10^{-3}$) [21]. Therefore, for our free-space measurements, any back reflection from the detector to the focusing lens and back to the detector is small ($<5 \cdot 10^{-5}$), resulting in a negligible contribution to the final systematic uncertainty. The SiTrap current readout is done by use of a high accuracy current-to-voltage amplifier (SiTrap amplifier) and a high accuracy voltmeter.

## 2.3 Pulsed versus CW measurement modes

For our free-space measurements we employed a narrow bandwidth continuous wave (CW) and a pulsed, mode-locked Ti:sapphire oscillator. When using the CW laser, we observed some fringing due to interference between the two window surfaces and the detector surface. The Ti:sapphire oscillator, with its short pulse duration and wide spectral bandwidth, eliminates any evidence of interference.

Measurements employing pulsed light may yield different responsivities than measurements made with CW light even though their average powers are the same due to detector nonlinearities at high peak incident light levels. Because our pulsed laser repetition rate was high ($\approx 76$ MHz), our average optical input power was low (< 10 µW) and the temporal response of our SiTrap and PM$_{mon}$ was slow (< 1 kHz), the systematic deviations between CW and pulsed mode measurements are negligible [22]. For an SPD at low count rates, the count rate depends linearly on the average input power, while at high count rates the dead time of the detector reduces the ratio of count rate to incident power. This is due to increased probability of photons arriving during the detector's dead time. This effect is also known as blocking loss [23]. For a weak CW laser, with photon arrival times according to a homogenous Poisson process, the blocking loss probability can be assumed linear with input count rate for $\mu_{CW}$<<1, which is defined as the probability of a photon arriving within the detector deadtime [23]. Similarly, for a weak pulsed laser (mean photon number per pulse $\mu_p$<<1), the blocking loss can also be assumed linear. At high count rates, the DE saturates, and the linear model is no longer adequate. However, in our case with maximum photon count rates of $10^6$ counts per second (cnt/s) [24][1], the deviation from the linear approximation is less than 0.01 %.

### 2.4 Afterpulsing characterization

We characterized the afterpulsing for two SPADs and one SNSPD at a wavelength of 851.8 nm with CW light. All photon detection events were time-tagged at count rates between 3000 cnt/s and 1.2·$10^6$ cnt/s, for a minimum of 30 seconds or at least $10^6$ detection events. For each dataset, we computed the number of counts per time bin of the sums of interarrival times between each detection and all subsequent ones, mapped the probability of subsequent detection events, which we then used to quantify afterpulsing or dark count rates [25]. The shape of the response can be seen in the plot of the interarrival time sums computed from the time tag data for detector NIST8103 (Figure 3(a); the peak at zero-delay is not shown). The signal is zero for times shorter than the dead time of the detector, of 52.29(20) ns. When the detector turns back on, a peak with an exponential decay is seen. Note, that for a homogeneous Poisson distribution we would expect to observe a flat response, *i.e.* the probability of detecting a photon at any given time is constant. We take the average of the number of counts beyond 500 ns as a baseline (solid red line in Figure 3(a)) and subtract it from the total measured signal to determine the total number of afterpulsing counts. The ratio of the remaining counts and the baseline signal is the afterpulsing probability, shown for three different detectors versus count rate in Figure 3(b). The inset in Figure 3(a) shows the of the sums of interarrival time bin counts for SNSPD PD9D where no afterpulsing is observed. Figure 3(b) shows the afterpulsing probability as a function of count rate for detectors NIST8103, V23173 and PD9D. The afterpulsing probability is defined as the ratio of the number of afterpulsing events and the detected photon counts. For NIST8103, the afterpulsing probability increases. This is a well-known and quantified effect [26]. It has been explained in terms of a "twilight" regime, where a photon absorption occurs while the SPAD voltage bias is returning to its full level but has not yet reached it. In that situation, the detector output is somewhat delayed in time. This effect becomes more pronounced as the average photon flux onto the device increases [25]. Detector V23173 shows a decreasing afterpulsing probability with increasing count rate. While this behavior is somewhat unusual, the two SPAD detector modules have different readout circuits, and some readout circuits are known to suppress twilight events in the output. This can lead to an apparent reduction in the afterpulsing probability as the probability of twilight events increases [25]. For detector PD9D the afterpulsing probability is seen to be negligible (Figure 3(b)).

---

[1] Here, we are referencing preferred notation for dimensionless units in the SI

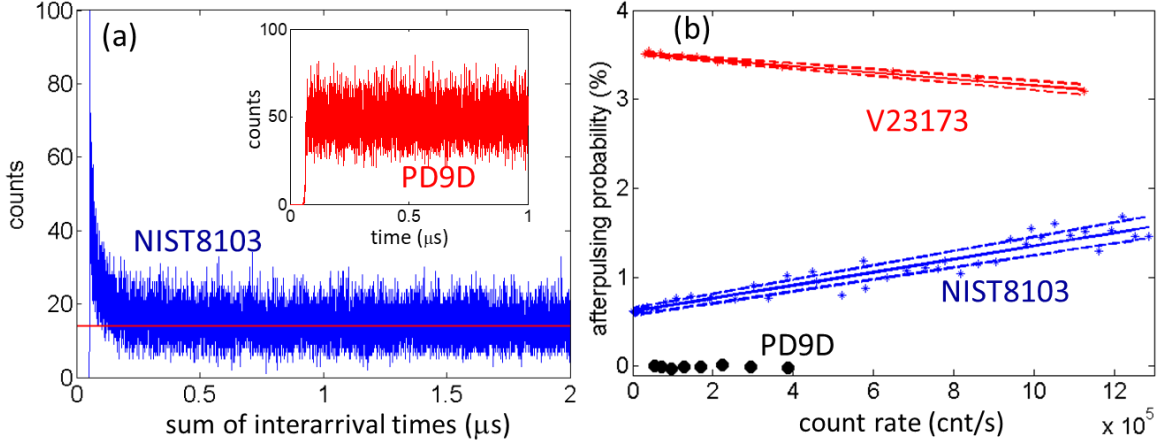

Figure 3: (a) Number of counts per time bin of the sums of interarrival times computed from the time-tag data from SPAD NIST8103 and SNSPD PD9D (inset) at count rates of ≈110000 cnt/s and ≈141000 cnt/s, respectively. The histogram bin size was set to the native resolution of the time tagger (156.25 ps). The time tagger's dead time was ≈ 5 ns. (b) Afterpulsing probability as a function of count rate for two SPADs and one SNSPD. The solid lines are linear fits. The dashed lines represent the 95 % confidence bounds.

### 2.5 Allan deviation of the laser sources

To test the stability of our lasers, we determined the relative Allan deviation [27] of the laser powers through the setup and the relative Allan deviation of the ratio of both powers measured at the DUT and at the $PM_{mon}$ locations. The relative Allan deviations are expressed as the percentage of the ratios between the Allan deviations and the average measured power and average measured ratio, respectively. We measured the power once every second and computed the relative Allan deviation of the laser power at $PM_{mon}$ as a function of averaging time, shown in Figure 4.

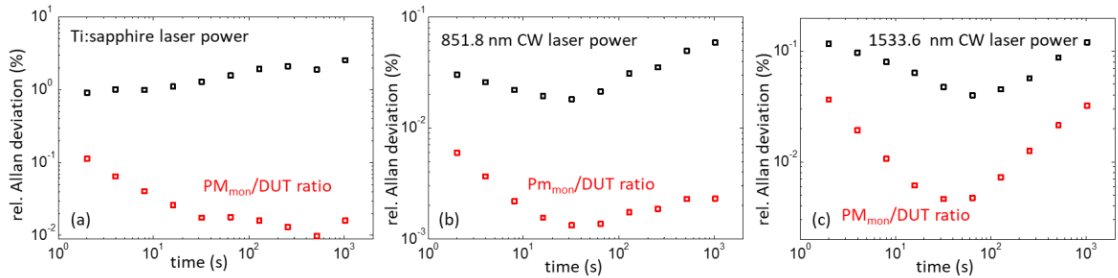

Figure 4: Relative Allan deviation of optical power as measured by $PM_{mon}$ for three different lasers (black squares) and relative Allan deviation of the ratio of the power at the DUT location and the $PM_{mon}$ (red squares). (a) Ti:sapphire laser. (b) 851.8 nm CW laser. (c) 1533.6 nm CW laser.

The DE measurements take an average of 25 s to complete. Therefore, we quote the laser power relative Allan deviation of about 1.2 %, 0.02 %, and 0.05 % at an averaging time of 25 s for the Ti:sapphire, 851.8 nm CW and 1533.6 nm CW laser, respectively. We also calculated the relative Allan deviation of the ratio between PM and $PM_{mon}$, which is seen to be significantly reduced relative to the raw laser power measurements. At an averaging time of 25 s, the relative Allan deviation of the ratio is estimated to be less than 0.02 %, 0.002 % and 0.005 % for the Ti:sapphire, 851.8 nm CW, and 1533.6 nm CW laser, respectively. While the observed laser power drift for the Ti:sapphire laser was not ideal, the ratio of the two power-meter readings was stable enough for our splitter/attenuator-based measurement protocol.

### 2.6 Polarization dependence

The DE of polarization-sensitive detectors, such as the SNSPD [28], was determined after maximizing the single-photon count rate at the detector using a manual fiber polarization controller added between VFA and the PM/DUT for the fiber-based measurements. The fiber polarization controller is fiber-loop-based and does not induce polarization-dependent loss. We reset and randomize the initial positions of the polarization controller before each measurement. We then maximize the single-photon count rate by adjusting the polarization controller.

### 2.7 Splicing versus fiber connector

For fiber-coupled detectors, we define the detection efficiency of an SPD as the system detection efficiency. The system detection efficiency (SDE) is the probability that a photon inside the optical fiber connected to the detector results in a discernable output signal. To transfer the optical power from the PM to the DUT, the fiber must be disconnected from the PM and connected to the DUT. For most practical purposes an FC/PC union connection would suffice to disconnect and connect the fibers. However, FC/PC union connections can pose a large uncertainty in the connector loss, since small misalignments between the fiber cores lead to losses at the connector, varying with each connector/connector combination and each breaking and mating of the connection. We have found connector losses ranging from negligible to ≈5 %. Alternately, one can fusion splice both fibers and achieve a lower connection-loss uncertainty. Below, we explore both fiber-to-DUT coupling methods.

### 2.8 Calibration chain

NIST calibration services provided our optical power-meter calibrations. With the above beam-splitter method and the calibration of the PM, we transferred the measurement scale to the single-photon domain. The calibration chain at NIST (Figure 5) ties the SPD calibrations to our optical fiber power meter and calibration services [19, 20, 29].

The primary standard for our calibrations is the Laser Optimized Cryogenic Radiometer (LOCR), which is tied to electrical standards [30]. With the LOCR, a pyroelectric transfer standard is calibrated to transfer the scale to the optical-fiber power-meter calibration service [19, 29], which then calibrates the PM and SiTrap used in this study. Our calibration service is used to measure the linearity of the $PM_{mon}$ and PM [20]. We utilize our spectral responsivity calibration service to measure the spectral uniformity of our SiTrap around 851 nm.

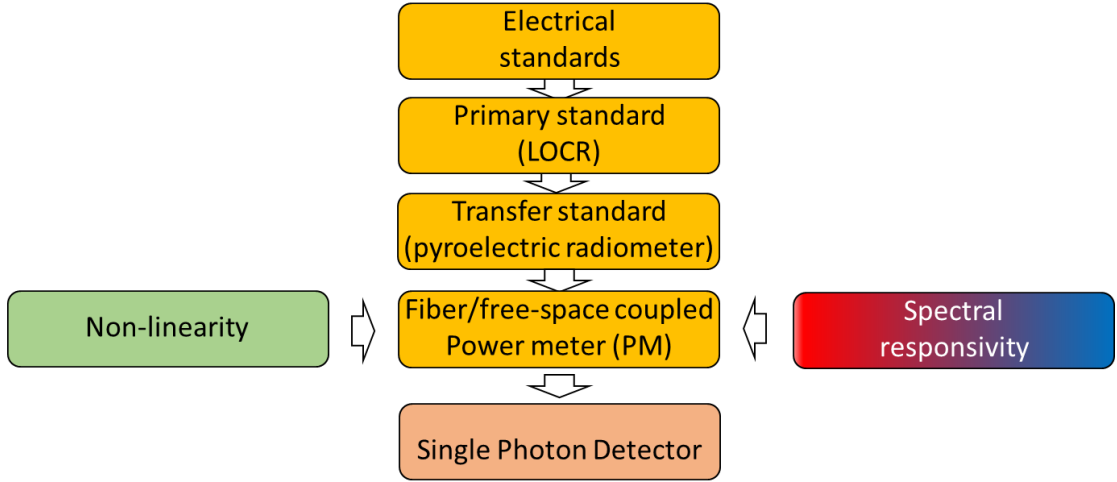

Figure 5: NIST calibration chain. We transfer the calibration of a PM using the beam-splitter method and by attenuating a laser beam inside an optical fiber. The PM is calibrated through a pyroelectric radiometer, which is calibrated by our laser-optimized cryogenic radiometer. Our services for calibration and spectral responsivity aid the transfer to optical powers compatible with single-photon detection rates at the DUT and correction for a specific wavelength of interest.

### 3. Estimation of uncertainties

The uncertainty budgets presented here are for DE at one count rate. Included are the uncertainty evaluations associated with the calibration of the SiTrap, $PM_{mon}$, PM, and the optical spectrum analyzer (OSA) calibration. The fiber-coupled measurements include an uncertainty associated with the fiber-end reflection-loss correction. The free-space uncertainty estimate includes the uncertainties associated with the SiTrap readout (transimpedance amplifier and voltmeter) and the free-space beam-collection efficiency.

To determine $DE$ at 1 cnt/s ($10^5$ cnt/s), we acquire the DE at many count rates. A linear fit is applied and by extrapolation (interpolation) we determine the mean DE value at 1 cnt/s ($10^5$ cnt/s). We calculated the squared uncertainty of the extrapolated (interpolated) DE by quadrature addition of the mean uncertainty established from uncertainties of all count rates and the uncertainty of the mean DE predicted by the linear fit.

#### 3.1 Measurement of detection efficiency at a given count rate

We use different measurement equations for the free-space and fiber-coupled measurements and the type of calibrated power meter. Our free-space measurements used the SiTrap, while our fiber-coupled measurement used PM(851.8 nm) and PM(1533.6 nm).

#### 3.1.1  Fiber-coupled

When using PM(851.8 nm) and PM(1533.6 nm), the *DE* measurement equation is:

$$DE = \frac{C^{\text{diff}}}{PM_{\text{mon}}^{\text{diff}}} \cdot \frac{h \cdot c}{\lambda_{\text{c}} \cdot \eta_{\text{f}}} \cdot \frac{1}{R_{\text{out/mon}}}, \qquad (1)$$

With

$$C^{\text{diff}} = \bar{C} - \bar{C}^{\text{drk}} - AP, \tag{2}$$

where $C^{\text{diff}}$ is the average photon count ($\bar{C}$) minus the sum of the average dark counts ($\bar{C}^{\text{drk}}$) and the afterpulsing counts ($AP$). We express $AP$ with a quadratic dependence on $\bar{C}$: $AP = (ap_0 + ap \cdot \bar{C}) \cdot \bar{C}$, where $ap$ and $ap_0$ are fitting parameters. $C^{\text{diff}}$ becomes

$$C^{\text{diff}} = \bar{C} - \bar{C}^{\text{drk}} - (ap_0 + ap \cdot \bar{C}) \cdot \bar{C} = (1 - ap_0) \cdot \bar{C} - \bar{C}^{\text{drk}} - ap \cdot \bar{C}^2. \tag{3}$$

$PM_{\text{mon}}^{\text{diff}}$ is the dark-current subtracted and calibrated optical power at PM$_{\text{mon}}$, and is expressed by:

$$PM_{\text{mon}}^{\text{diff}} = \frac{\overline{PM}_{\text{m}}^{\text{diff}}}{cal_{\text{nl}}^{PM\text{mon}}} \tag{4}$$

were $cal_{\text{nl}}^{PM\text{mon}}$ is the nonlinearity-correction factor for PM$_{\text{mon}}$, and $\overline{PM}_{\text{m}}^{\text{diff}}$ is the averaged difference between $n$ power readings and the average dark reading of PM$_{\text{mon}}$:

$$\overline{PM}_{\text{m}}^{\text{diff}} = \frac{1}{n}\sum_{i=1}^{n}\left(\left(PM_{\text{m}}^{\text{diff}}\right)_i + \overline{PM}_{\text{m/OSA}}^{\text{diff}}\right) = \frac{1}{n}\sum_{i=1}^{n}\left(PM_{\text{m}}^{\text{diff}}\right)_i + \overline{PM}_{\text{m/OSA}}^{\text{diff}}, \tag{5}$$

where $\left(PM_{\text{m}}^{\text{diff}}\right)_i$ represents the difference between the $i^{\text{th}}$ power reading and the average dark reading of PM$_{\text{mon}}$. Because the responsivity of our optical power meters depends on the input wavelength, we added $\overline{PM}_{\text{m/OSA}}^{\text{diff}}$ representing the power reading correction of PM$_{\text{mon}}$ due to the OSA calibration, *i.e.* the wavelength.

$$\overline{PM}_{\text{m/OSA}}^{\text{diff}} = \overline{PM}_{\text{OSA}}^{\text{diff}} \cdot b_\lambda \cdot \Delta\lambda_{\text{OSA}}. \tag{6}$$

The quantities $b_\lambda$ and $\overline{PM}_{\text{OSA}}^{\text{diff}}$ are scaling factors that adjust for wavelength and power level, respectively. Values for $b_\lambda$ are given in

Table 1. $\Delta\lambda_{\text{OSA}}$ is a correction for the OSA wavelength reading and is assumed to be constant.

The second term of eqn. (1) is an energy per photon factor consisting of Planck's constant ($h$), the speed of light ($c$), the mean wavelength of the photon source ($\lambda_c$) and $\eta_f$, which is the correction factor for the fiber-end-reflection when coupling into free-space. Our fiber-coupled SPADs use a lens to focus light onto the active area, hence there is no fiber-to-fiber junction. As a result, this is effectively a free-space coupling, thus $\eta_f$ is not required.

Finally, the denominator of the last term of eqn. (1) is the output-to-monitor ratio, given by:

$$R_{\text{out/mon}} = \frac{X}{Y} + R_{\text{out/mon}}^{\text{stab}}, \tag{7}$$

where

$$X = \frac{\overline{PM}^{\text{diff}}}{cal_{\text{nl}}^{PM} \cdot cal_{\text{abs}}^{PM}} \tag{8}$$

$$Y = \frac{\overline{PM}_{\text{mon}(R_{\text{out/mon}})}^{\text{diff}}}{cal_{\text{nl } R_{\text{out/mon}}}^{PM\text{mon}}} \tag{9}$$

and $R_{\text{out/mon}}^{\text{stab}}$ represents the correction in the output-to-monitor ratio. Furthermore, $cal_{\text{nl } R_{\text{out/mon}}}^{PM\text{mon}}$, $cal_{\text{nl}}^{PM}$ are the nonlinearity-correction factors for the ranges of $PM_{\text{mon}}$ during the $R_{\text{out/mon}}$ measurement and for the range of PM during the $R_{\text{out/mon}}$ measurement, respectively; $cal_{\text{abs}}^{PM}$ is the absolute calibration factor for PM.

The expression for $\overline{PM}^{\text{diff}}$ is

$$\overline{PM}^{\text{diff}} = \frac{1}{n_X}\sum_{i=1}^{n_X}\left(\left(PM^{\text{diff}}\right)_i + \overline{PM}_{\text{OSA}}^{\text{diff}}\right) = \frac{1}{n_X}\sum_{i=1}^{n_X}\left(PM^{\text{diff}}\right)_i + \overline{PM}_{\text{OSA}}^{\text{diff}}, \quad (10)$$

where $\left(PM^{\text{diff}}\right)_i$ is the difference between the $i^{\text{th}}$ power reading and the average dark reading of PM, and the quantity $\overline{PM}_{\text{OSA}}^{\text{diff}}$ is the correction of the power reading due to the OSA calibration, as defined in eqn. (5):

$$\overline{PM}_{\text{OSA}}^{\text{diff}} = \overline{PM}_{\text{OSA}}^{\text{diff}^*} \cdot b_\lambda \cdot \Delta\lambda_{\text{OSA}}, \quad (11)$$

where $\overline{PM}_{\text{OSA}}^{\text{diff}^*}$ is a scaling factor that adjusts for power level.

Similar to eqn. 10, we find for $\overline{PM}_{\text{mon}(R_{\text{out/mon}})}^{\text{diff}}$:

$$\begin{aligned}\overline{PM}_{\text{mon}(R_{\text{out/mon}})}^{\text{diff}} &= \frac{1}{n_Y}\sum_{i=1}^{n_Y}\left(\left(PM_{\text{mon}(R_{\text{out/mon}})}^{\text{diff}}\right)_i + \overline{PM}_{\text{mon}(R_{\text{out/mon}})/\text{OSA}}^{\text{diff}}\right) \\ &= \frac{1}{n_Y}\sum_{i=1}^{n_Y}\left(PM_{\text{mon}(R_{\text{out/mon}})}^{\text{diff}}\right)_i + \overline{PM}_{\text{mon}(R_{\text{out/mon}})/\text{OSA}}^{\text{diff}},\end{aligned} \quad (12)$$

where $\left(PM_{\text{mon}(R_{\text{out/mon}})}^{\text{diff}}\right)_i$ represents the difference between the $i^{\text{th}}$ power reading and the average dark reading of $PM_{\text{mon}}$ when measuring the output-to-monitor ratio, and the quantity $\overline{PM}_{\text{mon}(R_{\text{out/mon}})/\text{OSA}}^{\text{diff}}$ represents the correction in the power reading due to the OSA calibration:

$$\overline{PM}_{\text{mon}(R_{\text{out/mon}})/\text{OSA}}^{\text{diff}} = \overline{PM}_{\text{mon}(R_{\text{out/mon}})/\text{OSA}}^{\text{diff}^*} \cdot b_\lambda \cdot \Delta\lambda_{\text{OSA}}, \quad (13)$$

where $\overline{PM}_{\text{mon}(R_{\text{out/mon}})/\text{OSA}}^{\text{diff}^*}$ is a scaling factor that adjusts for power level.

Influence quantities, calibrations, resulting uncertainties, and their units are summarized in Tables 2 and 3. Based on the measurement eqn. (1), the relative combined standard uncertainty (from eqn. 12 and note 4 in 5.2.2 of the "Guide to the Expression of Uncertainty in Measurement" (GUM) [31]) is:

$$\frac{u(DE)}{DE} = \sqrt{\left(\frac{u(C^{\text{diff}})}{C^{\text{diff}}}\right)^2 + \left(\frac{u(PM_{\text{mon}}^{\text{diff}})}{PM_{\text{mon}}^{\text{diff}}}\right)^2 + \left(\frac{u(\lambda_c)}{\lambda_c}\right)^2 + \left(\frac{u(\eta_f)}{\eta_f}\right)^2 + \left(\frac{u(R_{\text{out/mon}})}{R_{\text{out/mon}}}\right)^2 - 2\frac{u(C^{\text{diff}}, PM_{\text{mon}}^{\text{diff}})}{C^{\text{diff}} \cdot PM_{\text{mon}}^{\text{diff}}}}, \quad (14)$$

where

$$u^2(C^{\text{diff}}) = (1 - ap_0 - 2 \cdot ap \cdot \bar{C})^2 u^2(\bar{C}) + (-1)^2 u^2(\bar{C}^{\text{drk}}) + (-\bar{C})^2 u^2(ap_0) + \\ (-\bar{C}^2)^2 u^2(ap) + 2(-\bar{C}^2)(-\bar{C})u(ap_0, ap), \quad (15)$$

$$\left(\frac{u(PM_{\text{mon}}^{\text{diff}})}{PM_{\text{mon}}^{\text{diff}}}\right)^2 = \left(\frac{u(\overline{PM}_{\text{m}}^{\text{diff}})}{\overline{PM}_{\text{m}}^{\text{diff}}}\right)^2 + \left(\frac{u(cal_{\text{nl}}^{PM\text{mon}})}{cal_{\text{nl}}^{PM\text{mon}}}\right)^2, \tag{16}$$

$$\left(\frac{u(R_{\text{out/mon}})}{R_{\text{out/mon}}}\right)^2 = \frac{1}{(R_{\text{out/mon}})^2}\left[u^2(R_{\text{out/mon}}^{\text{stab}}) + \left(\frac{1}{Y}\right)^2 u^2(X) + \left(\frac{-X}{Y^2}\right)^2 u^2(Y) + 2\left(\frac{1}{Y}\right)\left(\frac{-X}{Y^2}\right)u(X,Y)\right], \tag{17}$$

$$u^2(X) = X^2\left[\left(\frac{u(\overline{PM}^{\text{diff}})}{\overline{PM}^{\text{diff}}}\right)^2 + \left(\frac{u(cal_{\text{nl}}^{PM})}{cal_{\text{nl}}^{PM}}\right)^2 + \left(\frac{u(cal_{\text{abs}}^{PM})}{cal_{\text{abs}}^{PM}}\right)^2\right], \tag{18}$$

and

$$u^2(Y) = Y^2\left(\frac{u\left(\overline{PM}_{\text{mon}(R_{\text{out/mon}})}^{\text{diff}}\right)}{\overline{PM}_{\text{mon}(R_{\text{out/mon}})}^{\text{diff}}}\right)^2 + \left(\frac{u\left(cal_{\text{nl } R_{\text{out/mon}}}^{PM\text{mon}}\right)}{cal_{\text{nl } R_{\text{out/mon}}}^{PM\text{mon}}}\right)^2. \tag{19}$$

The data shown in Figure 2 are used to estimate $u(R_{\text{out/mon}}^{\text{stab}})$. Two covariances, $u(C^{\text{diff}}, PM_{\text{mon}}^{\text{diff}})$ and $u(X,Y)$, are known to exist, mainly due to simultaneous measurement. We characterized the covariances and found that both are positive. We omit both positive correlations in our uncertainty estimation, therefore overestimating the final uncertainty.

The uncertainties in the power reading and dark reading are not separately estimated, so the uncertainty is based on the difference between the two readings. The squared uncertainty of $\overline{PM}_{\text{m}}^{\text{diff}}$ is

$$u^2\left(\overline{PM}_{\text{m}}^{\text{diff}}\right) = \frac{1}{n}s^2 + u^2\left(\overline{PM}_{\text{m/OSA}}^{\text{diff}}\right) \tag{20}$$

$$u^2\left(\overline{PM}_{\text{m/OSA}}^{\text{diff}}\right) = \left(\overline{PM}_{\text{m/OSA}}^{\text{diff}}\right)^2\left[\left(\frac{u(\overline{PM}_{\text{OSA}}^{\text{diff}})}{\overline{PM}_{\text{OSA}}^{\text{diff}}}\right)^2 + \left(\frac{u(b_\lambda)}{b_\lambda}\right)^2\right], \tag{21}$$

where $s^2$ represents the sample variance of $n$ measurements of $\left(PM_{\text{m}}^{\text{diff}}\right)_i$, and $u(b_\lambda)$ represents the uncertainty of $b_\lambda$.

The squared uncertainty of $\overline{PM}^{\text{diff}}$ is

$$u^2\left(\overline{PM}^{\text{diff}}\right) = \frac{1}{n_X}s_X^2 + u^2\left(\overline{PM}_{OSA}^{\text{diff}}\right) \tag{22}$$

$$u^2\left(\overline{PM}_{\text{OSA}}^{\text{diff}}\right) = \left(\overline{PM}_{\text{OSA}}^{\text{diff}}\right)^2\left[\left(\frac{u(\overline{PM}_{\text{OSA}}^{\text{diff}^*})}{\overline{PM}_{\text{OSA}}^{\text{diff}^*}}\right)^2 + \left(\frac{u(b_\lambda)}{b_\lambda}\right)^2\right]. \tag{23}$$

The term $s_X^2$ represents the sample variance of $n_X$ measurements of $\left(PM^{\text{diff}}\right)_i$. A similar equation is used to estimate $u^2\left(\overline{PM}_{\text{mon}(R_{\text{out/mon}})}^{\text{diff}}\right)$,

$$u^2\left(\overline{PM}_{\text{mon}(R_{\text{out/mon}})}^{\text{diff}}\right) = \frac{1}{n_Y}s_Y^2 + u^2\left(\overline{PM}_{\text{mon}(R_{\text{out/mon}})/\text{OSA}}^{\text{diff}}\right) \tag{24}$$

$$u^2\left(\overline{PM}_{\text{mon}(R_{\text{out/mon}})/\text{OSA}}^{\text{diff}}\right) =$$
$$\left(\overline{PM}_{\text{mon}(R_{\text{out/mon}})/\text{OSA}}^{\text{diff}}\right)^2 \left[\left(\frac{u(\overline{PM}_{\text{mon}(R_{\text{out/mon}})/\text{OSA}}^{\text{diff}^*})}{\overline{PM}_{\text{mon}(R_{\text{out/mon}})/\text{OSA}}^{\text{diff}^*}}\right)^2 + \left(\frac{u(b_\lambda)}{b_\lambda}\right)^2\right] \quad (25)$$

where $s_Y^2$ is the sample variance of $n_Y$ measurements of $\left(PM_{\text{mon}(R_{\text{out/mon}})}^{\text{diff}}\right)_i$.

### 3.1.2 Free space

For the free-space based measurements, where no fiber-end-reflection correction is applied, the measurement equation is

$$DE_{\text{FSM}} = DE_{\text{free}} + DE_{\text{reflect}} + DE_{\text{collect}} + DE_{\text{align}}, \quad (26)$$

where

$$DE_{\text{free}} = \frac{C^{\text{diff}}}{PM_{\text{mon}}^{\text{diff}}} \cdot \frac{h \cdot c}{\lambda_c} \cdot \frac{1}{RV_{\text{out/mon}}} . \quad (27)$$

Reflection, collection, and alignment sources of variability in eqn. (26) are depicted by $DE_{\text{reflect}}$, $DE_{\text{collect}}$, and $DE_{\text{align}}$, respectively. $C^{\text{diff}}$ is defined in eqn. (3), $PM_{\text{mon}}^{\text{diff}}$ is defined in eqn. (4), $h$, $c$ and $\lambda_c$ are defined above. For the free-space measurement, we use the SiTrap as the calibrated power meter. The output-to-monitor ratio (denominator of the third term in eqn. (27)) becomes:

$$RV_{\text{out/mon}} = \frac{\overline{V}_{\text{trap}} - \overline{V}_{\text{trap}}^{\text{drk}}}{PM_{\text{mon}(RV_{\text{out/mon}})} - PM_{\text{mon}(RV_{\text{out/mon}})}^{\text{drk}}} \cdot \frac{cal_{\text{nl } RV_{\text{out/mon}}}^{PM\text{mon}}}{\overline{R} \cdot g} + R_{\text{out/mon}}^{\text{stab}}$$
$$= \frac{\overline{V}_{\text{trap}} - \overline{V}_{\text{trap}}^{\text{drk}}}{\overline{PM}_{\text{mon}(RV_{\text{out/mon}})}^{\text{diff}}} \cdot \frac{cal_{\text{nl } RV_{\text{out/mon}}}^{PM\text{mon}}}{\overline{R} \cdot g} + R_{\text{out/mon}}^{\text{stab}}, \quad (28)$$

where $\overline{V}_{\text{trap}}$ and $\overline{V}_{\text{trap}}^{\text{drk}}$ are the bright and dark voltage readings from the SiTrap, and SiTrap amplifier unit, respectively. The quantities $\overline{PM}_{\text{mon}(RV_{\text{out/mon}})}^{\text{diff}}$ (eqn. (12)), $R_{\text{out/mon}}^{\text{stab}}$ and $cal_{\text{nl } RV_{\text{out/mon}}}^{PM\text{mon}}$ are defined as in the fiber-coupled experiment. $\overline{R}$ is the responsivity of the SiTrap and $g$ is transimpedance amplifier gain. Note that we use the same amplifier gain setting and measurement range on the voltmeter when measuring $\overline{V}_{\text{trap}}$ and $\overline{V}_{\text{trap}}^{\text{drk}}$.

We rewrite eqn. (28) as

$$RV_{\text{out/mon}} = \frac{W}{Y} + R_{\text{out/mon}}^{\text{stab}}, \quad (29)$$

where $Y$ is defined in eqn. (9) and

$$W = \frac{\overline{V}_{\text{trap}} - \overline{V}_{\text{trap}}^{\text{drk}}}{\overline{R} \cdot g} . \quad (30)$$

The bright and dark voltage readings are defined as

$$\bar{V}_{\text{trap}} = \frac{1}{n_t}\sum_{i=1}^{n_t}\left((V_{\text{trap}})_i + \bar{V}_{\text{trap/cal}}\right) = \frac{1}{n_t}\sum_{i=1}^{n_t}(V_{\text{trap}})_i + \bar{V}_{\text{trap/cal}}, \tag{31}$$

and

$$\bar{V}_{\text{trap}}^{\text{drk}} = \frac{1}{n_{td}}\sum_{j=1}^{n_{td}}\left((V_{\text{trap}}^{\text{drk}})_j + \bar{V}_{\text{trap/cal}}\right) = \frac{1}{n_{td}}\sum_{j=1}^{n_{td}}(V_{\text{trap}}^{\text{drk}})_j + \bar{V}_{\text{trap/cal}}. \tag{32}$$

$\bar{V}_{\text{trap/cal}}$ denotes the correction in the voltage reading due to the voltmeter calibration. The responsivity ($\bar{R}$) of the trap is:

$$\bar{R} = \bar{R}_{\text{cal}} + \bar{R}_{\text{OSA}}, \tag{33}$$

The quantity $\bar{R}_{\text{cal}}$ is the absolute SiTrap responsivity. The quantity, $\bar{R}_{\text{OSA}}$, represents the correction of the responsivity due to the OSA calibration:

$$\bar{R}_{\text{OSA}} = \bar{R}_{\text{OSA}}^{*} \cdot b_\lambda \cdot \Delta\lambda_{\text{OSA}}, \tag{34}$$

where $\bar{R}_{\text{OSA}}^{*}$ is a scaling factor that adjusts for trap responsivity.

Uncertainties associated with each influence quantity are summarized in Tables 2 and 3 are consolidated for $C^{\text{diff}}$, $PM_{\text{mon}}^{\text{diff}}$, $RV_{\text{out/mon}}$. Uncertainties for additional influence quantities for $DE_{\text{FSM}}$, reflection, collection, and alignment, are also shown in Tables 2 and 3.

For the free-space case, based on measurement eqn. (26), the squared combined standard uncertainty is

$$u^2(DE_{\text{FSM}}) = u^2(DE_{\text{free}}) + u^2(DE_{\text{reflect}}) + u^2(DE_{\text{collect}}) + u^2(DE_{\text{align}}), \tag{35}$$

and the relative combined standard uncertainty is

$$\frac{u(DE_{\text{FSM}})}{DE_{\text{FSM}}} = \sqrt{\frac{u^2(DE_{\text{free}})}{DE_{\text{FSM}}^2} + \frac{u^2(DE_{\text{reflect}})}{DE_{\text{FSM}}^2} + \frac{u^2(DE_{\text{collect}})}{DE_{\text{FSM}}^2} + \frac{u^2(DE_{\text{align}})}{DE_{\text{FSM}}^2}}. \tag{36}$$

Uncertainties due to reflection, collection, and alignment are determined from characterization measurements of the free-space coupled detector itself and the focused free-space beam are discussed below. The squared uncertainty of $DE_{\text{free}}$ is computed by applying:

$$u^2(DE_{\text{free}}) = DE_{\text{free}}^2\left[\left(\frac{u(C^{\text{diff}})}{C^{\text{diff}}}\right)^2 + \left(\frac{u(PM_{\text{mon}}^{\text{diff}})}{PM_{\text{mon}}^{\text{diff}}}\right)^2 + \left(\frac{u(\lambda_c)}{\lambda_c}\right)^2 + \left(\frac{u(RV_{\text{out/mon}})}{RV_{\text{out/mon}}}\right)^2\right], \tag{37}$$

where $u(C^{\text{diff}})$ is defined in eqn. (15) and $\left(\frac{u(PM_{\text{mon}}^{\text{diff}})}{PM_{\text{mon}}^{\text{diff}}}\right)^2$ is defined in eqn. (16). $\left(\frac{u(\lambda_c)}{\lambda_c}\right)^2$ can be calculated using the OSA calibration uncertainty and source wavelength. In addition,

$$\left(\frac{u(RV_{\text{out/mon}})}{RV_{\text{out/mon}}}\right)^2 = \frac{1}{(RV_{\text{out/mon}})^2}\left[u^2(R_{\text{out/mon}}^{\text{stab}}) + \left(\frac{-W}{Y^2}\right)^2 u^2(Y) + \left(\frac{1}{Y}\right)^2 u^2(W) + 2\left(\frac{1}{Y}\right)\left(\frac{-W}{Y^2}\right)u(W,Y)\right], \tag{38}$$

where $u(Y)$ is defined as in eqn. (19). The covariances between $C^{\text{diff}}$ and $PM_{\text{mon}}^{\text{diff}}$, and between $W$ and $Y$, are positive and will be ignored in the uncertainty calculation, resulting in a slightly conservative uncertainty estimate.

Applying the law of propagation of uncertainty to the expression for $W$ produces:

$$u^2(W) = \left(\frac{1}{\bar{R}\cdot g}\right)^2 u^2(\bar{V}_{\text{trap}}) + \left(\frac{-1}{\bar{R}\cdot g}\right)^2 u^2(\bar{V}_{\text{trap}}^{\text{drk}}) + \left(\frac{\bar{V}_{\text{trap}}-\bar{V}_{\text{trap}}^{\text{drk}}}{-\bar{R}^2\cdot g}\right)^2 u^2(\bar{R}) \left(\frac{\bar{V}_{\text{trap}}-\bar{V}_{\text{trap}}^{\text{drk}}}{-\bar{R}\cdot g^2}\right)^2 u^2(g), \quad (39)$$

which can be re-written as

$$\left(\frac{u(W)}{W}\right)^2 = \left(\frac{u(\bar{V}_{\text{trap}})}{\bar{V}_{\text{trap}}-\bar{V}_{\text{trap}}^{\text{drk}}}\right)^2 + \left(\frac{u(\bar{V}_{\text{trap}}^{\text{drk}})}{\bar{V}_{\text{trap}}-\bar{V}_{\text{trap}}^{\text{drk}}}\right)^2 + \left(\frac{u(\bar{R})}{\bar{R}}\right)^2 + \left(\frac{u(g)}{g}\right)^2, \quad (40)$$

where $u(g)$ is the uncertainty associated with the SiTrap amplifier calibration, and:

$$u^2(\bar{V}_{\text{trap}}) = \frac{1}{n_t}s_t^2 + u^2(\bar{V}_{\text{trap/cal}}), \quad (41)$$

$$u^2(\bar{V}_{\text{trap}}^{\text{drk}}) = \frac{1}{n_{td}}s_{td}^2 + u^2(\bar{V}_{\text{trap/cal}}), \quad (42)$$

The quantities, $s_t^2$ and $s_{td}^2$ are sample variances, and

$$u^2(\bar{R}) = u^2(\bar{R}_{\text{cal}}) + u^2(\bar{R}_{\text{OSA}}) \quad (43)$$

$$u^2(\bar{R}_{\text{OSA}}) = (\bar{R}_{\text{OSA}})^2 \left[\left(\frac{u(\bar{R}_{\text{OSA}}^*)}{\bar{R}_{\text{OSA}}^*}\right)^2 + \left(\frac{u(b_\lambda)}{b_\lambda}\right)^2\right] \quad (44)$$

$u(\bar{V}_{\text{trap/cal}})$ and $u(\bar{R}_{\text{cal}})$ are the uncertainties associated with the voltmeter calibration and the absolute SiTrap responsivity calibration, respectively.

Table 1: Values of $b_\lambda$ for PM$_{\text{mon}}$, PM(851.8 nm), PM(1533.6 nm) and SiTrap at the measurement wavelengths and their associated uncertainties ($k=1$).

|  | $b_\lambda$ (nm$^{-1}$) | | | |
| --- | --- | --- | --- | --- |
| λ | PM$_{\text{mon}}$ | PM(851.8 nm) | PM(1533.6 nm) | SiTrap (851.8 nm) |
| 851.8 nm | -0.01028(4) | -0.00244(5) | --- | 0.000808(8) |
| 1533.6 nm | 0.00022(1) | --- | -0.000008(2) | --- |

Table 2: Component uncertainties ($k=1$) for measurands to determine the DE at a single count rate for fiber-coupled and free-space measurements.

| measurand | fiber coupled | free space | uncertainty | type | unit | component type |
|---|---|---|---|---|---|---|
| $C^{\text{diff}}$ | ✓ | ✓ | $u(\bar{C}) = \sqrt{C/N}$ | A | unitless | shot noise |
|  | ✓ | ✓ | $u(\bar{C}^{\text{drk}}) = \sqrt{C/N}$ | A | unitless | shot noise |
|  | ✓ | ✓ | $u(ap), u(ap_0),$ $u(ap, ap_0)$ | B | unitless | determined by fitting the eqn. (3) model to the observed data |
| $PM_{\text{mon}}^{\text{diff}}$ | ✓ | ✓ | $\dfrac{u(cal_{\text{nl}}^{PM\text{mon}})}{cal_{\text{nl}}^{PM\text{mon}}}$ | B | % | nonlinearity correction |
| $PM_{\text{mon}(R_{\text{out/mon}})}^{\text{diff}}$ | ✓ | ✓ | $\dfrac{u(cal_{\text{nl } R_{\text{out/mon}}}^{PM\text{mon}})}{cal_{\text{nl } R_{\text{out/mon}}}^{PM\text{mon}}}$ | B | % | nonlinearity correction |
| $PM^{\text{diff}}$ | ✓ |  | $\dfrac{u(cal_{\text{nl}}^{PM})}{cal_{\text{nl}}^{PM}}$ | B | % | nonlinearity correction |
| $PM^{\text{diff}}$ | ✓ |  | $\dfrac{u(cal_{\text{abs}}^{PM})}{cal_{\text{abs}}^{PM}}$ | B | % | absolute calibration |
| $R$ |  | ✓ | $\dfrac{u(\bar{R}_{cal})}{\bar{R}_{cal}}$ | B | % | absolute calibration |
| $g$ |  | ✓ | $\dfrac{u(g)}{g}$ | B | % | amplifier gain calibration |
| $V_{\text{trap}}, V_{\text{trap}}^{\text{drk}}$ |  | ✓ | $\dfrac{u(\bar{V}_{\text{trap}/cal})}{\bar{V}_{\text{trap}/cal}}$ | B | % | voltmeter calibration |
| $\lambda_c$ | ✓ | ✓ | $u(\lambda_c)$ | B | m | OSA calibration |
| $\eta_f$ | (✓) |  | $u(\eta_f)$ | B | unitless | fiber-end-reflection |
| $DE$ |  | ✓ | $\dfrac{u(DE_{\text{reflect}})}{DE_{\text{reflect}}}$ | B | % | SiTrap-focusing lens reflection |
|  |  | ✓ | $\dfrac{u(DE_{\text{collect}})}{DE_{\text{collect}}}$ | B | % | Beam collection |
|  |  | ✓ | $\dfrac{u(DE_{\text{align}})}{DE_{\text{align}}}$ | B | % | Free-space beam alignment |

Table 3: Summary of systematic standard uncertainties ($k$=1).

| | fiber-coupled | | | | free-space | |
|---|---|---|---|---|---|---|
| wavelength | 851 nm | | 1533.6 nm | | 851 nm | |
| item | value | unit | value | unit | value | unit |
| $\dfrac{u(cal_{nl}^{PMmon})}{cal_{nl}^{PMmon}}$ | 0.14 | % | 0.04 | % | 0.14 | % |
| $\dfrac{u(cal_{nl\,R_{out/mon}}^{PMmon})}{cal_{nl\,R_{out/mon}}^{PMmon}}$ | 0.14 | % | 0.04 | % | 0.14 | % |
| $\dfrac{u(R_{out/mon}^{stab})}{R_{out/mon}^{stab}}$ | 0.20 | % | 0.05 | % | 0.20 | % |
| $\dfrac{u(cal_{abs}^{PM})}{cal_{abs}^{PM}}$ | 0.22 | % | 0.19 | % | --- | --- |
| $\dfrac{u(cal_{nl}^{PM})}{cal_{nl}^{PM}}$ | 0.10 | % | 0.05 | % | --- | --- |
| $\dfrac{u(\bar{R}_{cal})}{\bar{R}_{cal}}$ | --- | --- | --- | --- | 0.22 | % |
| $\dfrac{u(g)}{g}$ | --- | --- | --- | --- | 0.01 | % |
| $\dfrac{u(\bar{V}_{trap/cal})}{\bar{V}_{trap/cal}}$ | --- | --- | --- | --- | 0.01 | % |
| $u(OSA)$ | $10^{-10}$ | m | $10^{-10}$ | m | $10^{-10}$ | m |
| $u(\eta_f)$ | $10^{-3}$ | --- | $10^{-3}$ | --- | $10^{-3}$ | --- |
| $\dfrac{u(DE_{reflect})}{DE_{reflect}}$ | --- | --- | --- | --- | $5 \cdot 10^{-3}$ | % |
| $\dfrac{u(DE_{collect})}{DE_{collect}}$ | --- | --- | --- | --- | 0.10 | % |
| $\dfrac{u(DE_{align})}{DE_{align}}$ | --- | --- | --- | --- | 0.50 | % |

## 4. Free-space characterization of a Si SPAD near 851 nm

Here, we present our results characterizing a free-space Si-SPAD at a wavelength near 851 nm and count rates of 1 cnt/s and $10^5$ cnt/s.

### 4.1 Experimental Setup

The experimental setup (Figure 6) begins with a fiber-coupled CW laser or Ti:sapphire oscillator coupled into an optical fiber at a typical power of ≈5 mW. The light is coupled to a free-space polarization controller consisting of a half-wave/quarter-wave/half-wave-plate combination. We use the polarization controller to set the polarization at the DUT and for measuring the DUT's polarization sensitivity. The polarization is also constantly monitored with a free-space polarization analyzer.

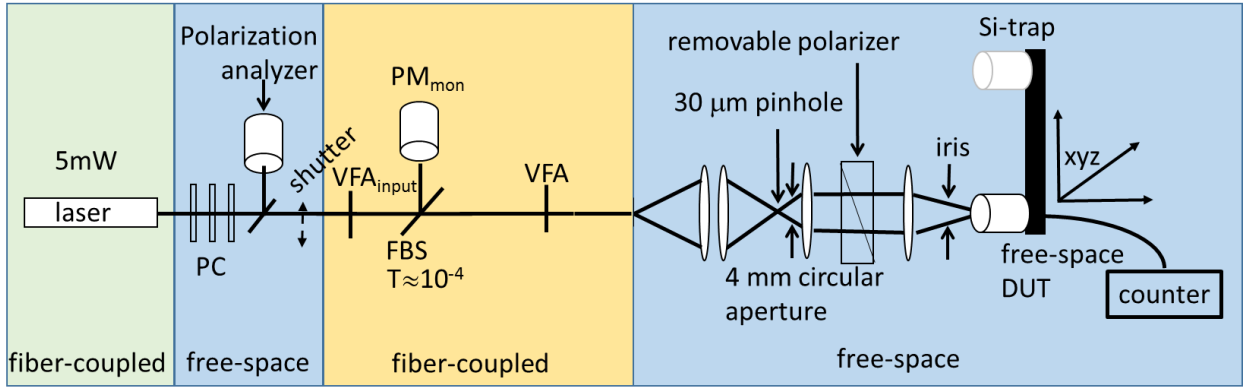

Figure 6: Experimental Setup: Light from free-space is coupled into a fiber, or a fiber laser is used. The light from the fiber is coupled into a free-space polarization controller (PC) and polarization analysis setup, after which the light is coupled into fiber and sent through a variable fiber attenuator (VFA$_{input}$). The high-power output of the 1:$10^4$ fiber beam splitter (FBS) is connected to a monitor power meter. The low-power output light is sent through VFA and into free-space collimation and focusing optics. An *xyz*-stage allows switching between the DUT and SiTrap. The single photons detected by the DUT are recorded by a counter. A removable polarizer is used to set the polarization.

We use a controllable shutter, serving as a mechanical switch to block or unblock the light entering the calibration setup. After the shutter, the light is coupled into an optical fiber that is multi-mode at 851 nm. This matches the optical fiber of the VFA$_{input}$. The output of VFA$_{input}$ is connected to a single-mode optical fiber for 851 nm and directed to a single-mode FBS. The nominal splitting ratio of the FBS is 1:$10^4$, where the high-power output port is connected to PM$_{mon}$. The low-power output port of the FBS is connected to VFA, after which the light is coupled to the free-space section of the setup, where it is collimated and re-focused onto a 30 μm diameter pinhole to clean up the spatial mode and remove Airy rings originating from the fiber. Collimation and further beam cleanup with a 4 mm circular aperture block the Airy rings originating from the pinhole. The beam is then focused with a 79 mm focal length aspheric lens. We chose a long-focal length lens and an additional variable iris after the lens to minimize the possibility of any back-reflections from the DUT reflecting back onto the DUT. In the focal plane, the beam 1/$e^2$-beam diameter is ≈20 μm. The SiTrap and DUT are co-mounted on an *xyz*-translation stage. Thus, the free-space optical beam is stationary, while the DUT and SiTrap can be moved into the optical beam. To set the polarization at the DUT, we temporarily place a polarizer into the free-space focusing path to act as a polarization analyzer. With the free-space polarization controller we can set the polarization at the DUT. After the polarization is set, we remove the polarization analyzer from the setup.

Photographs of the setup (Figure 7) show the free-space polarization controller and analyzer with the VFAs and the counter in the background (left), the free-space beam-shaping and focusing setup (center), and the *xyz*-stage to position the DUT and SiTrap for the free-space measurements (right). Both, the beam-shaping setup and *xyz*-stage setup are covered by black cardboard and cloth to minimize stray light reaching the DUT. Note that for our fiber-coupled measurements, we use the same experimental setup, except for the free-space beam-shaping and focusing optics.

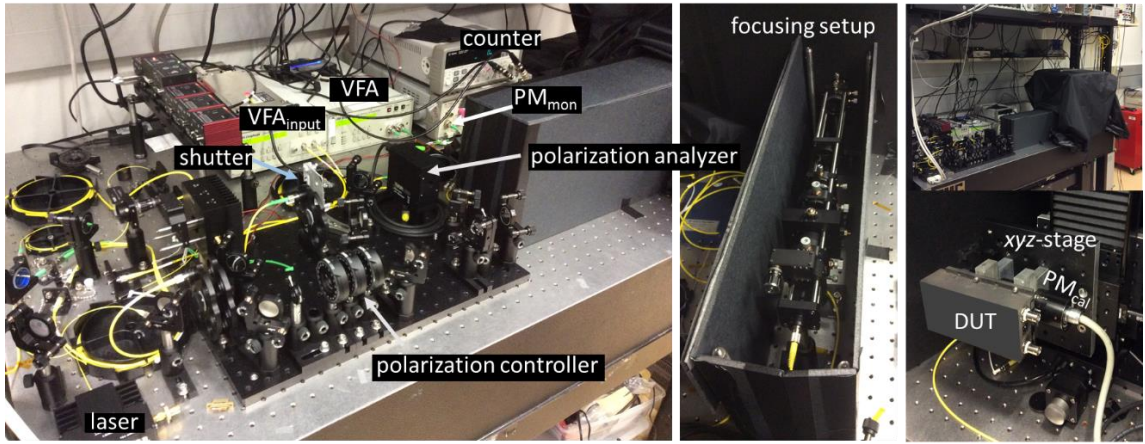

Figure 7: Photos of the experimental setup.

### 4.2 Free-space beam geometry

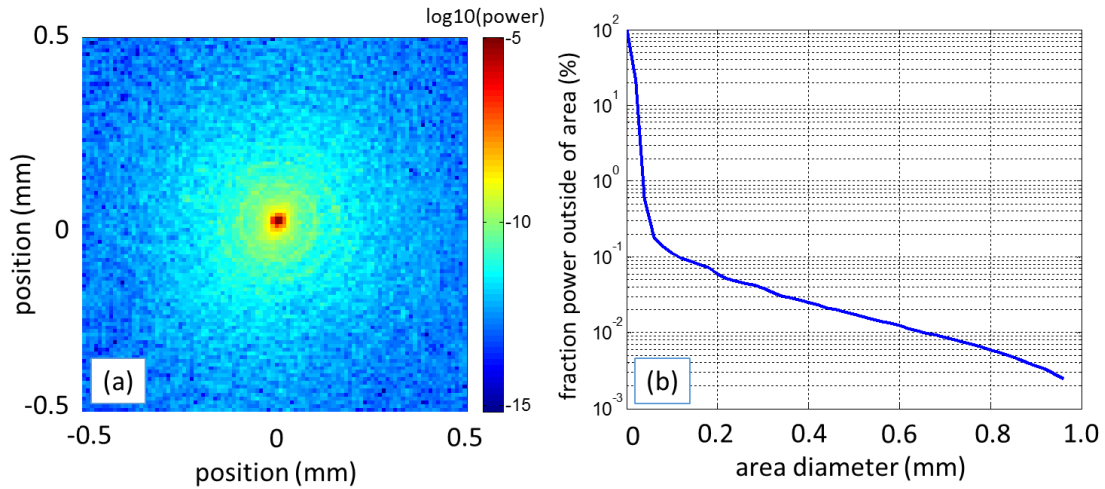

Figure 8: (a) Spatial beam scan at the focal position with a 10 μm pinhole on a logarithmic scale (b) Fraction of light intensity outside of a circular area centered on the peak light intensity.

We measured the spatial intensity distribution of the focused laser beam after beam-shaping at the location of the DUT. To quantify the collection efficiency of the light hitting the SPAD we used the *xyz*-stage to scan a 10 μm pinhole across the beam in 10 μm steps. The measured spatial intensity distribution (Figure 8 (a)) shows weak Airy rings around the main beam spot. The peak intensity of these rings is at least 5 orders of magnitude lower than the central beam intensity, and they result from the 4 mm aperture in the beam path before the collimating lens. Figure 8 (b) shows the fraction of the light intensity outside a circular area with diameter *D*. At a diameter of ≈150 μm, the fraction of the light outside this area drops below 0.1 %. At 0.2 mm, this ratio has dropped to ≈0.06 %.

### 4.3 SPAD spatial uniformity

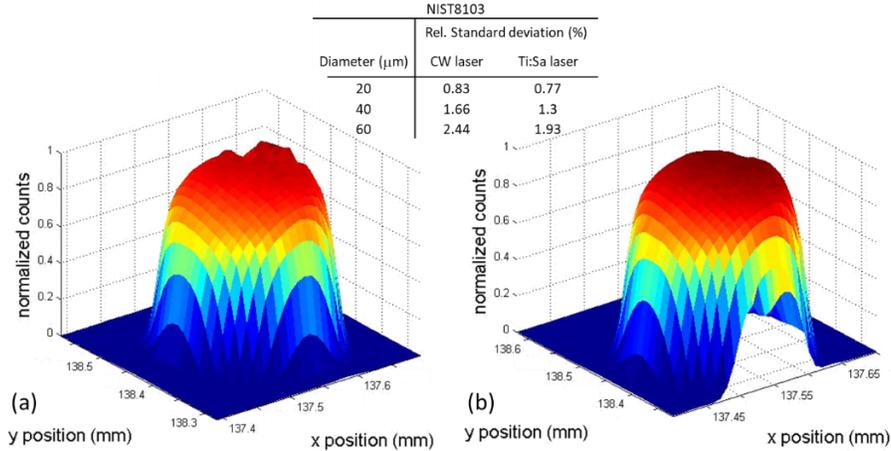

Figure 9: Spatial response scans of the NIST8103 SPAD using (a) a CW laser source and (b) a Ti:sapphire oscillator with short pulse duration. Standard deviations of response for the central region for a given diameter are indicated in the inset table: calculated standard deviations of counts within a circular area diameter for the NIST8103 SPAD. For the CW laser, the standard deviations of DUT responsivity for area diameters ranging from 20 μm to 60 μm are larger than for the Ti:sapphire oscillator.

Spatial scans were made of the NIST8103 SPAD by use of the CW and pulsed lasers. Scans made with the CW laser (Figure 9(a)) show interference fringes with a contrast of nearly 10 %, which most likely originate from reflections between the two window surfaces and the detector. With the pulsed laser, no fringes can be observed (Figure 9(b)). To quantify response uniformity, we calculated the standard deviation of the response for circular areas with three different diameters centered on the SPAD (table inset in Figure 9). The nonuniformity is slightly larger in the CW case due to the interference fringes, which are most prominent at the edge of the active area. We determined a relative slope of 0.1 %/μm in the detected count rate across the central portion of the active area. Since our setup allows us to find the center of the SPAD with repeatability better than 5 μm, we can estimate the uncertainty in the detector response due to detector alignment to be 0.5 %.

### 4.4 Free-space SPAD 851 nm calibration

Figure 10 depicts the measured DEs as a function of count rate for the NIST8103 SPAD. Figure 10(a) and Figure 10(b) show the results of DEs versus detector count rate with the CW laser, plotted on a linear scale and a $\log_{10}$-scale, respectively. Figure 10(c) and Figure 10(d) show the DEs with the Ti:sapphire laser. The half-width of the error bar for each DE measurement is the associated standard uncertainty of the measurement. At low count rates, the spread in the data and the uncertainty are increased. Therefore, we ensured that at each $VFA_{input}$ setting we acquired at least $10^6$ photon detection events. For example, at a count rate of 3000 cnt/s, we performed at least 14 measurements of about 25 s each. The average DE at each attenuator setting is plotted versus average count rate at each attenuator setting and represented by the red circles in Figure 10(b) and Figure 10(d). In both cases, *DE* appears to be flat at sufficiently low count rates within the statistical uncertainty. However, at larger count rates, *DE* decreases as the count rate increases. The decreasing efficiency with higher count rate, visible in the linear-scale plots is due to the blocking loss related to the SPAD's dead time. We fit a linear model for *DE* as a function of count rate. Based on the estimated intercept and slope parameters, we extrapolated to determine the DE of the SPAD at 1 cnt/s and interpolated to determine the DE of the SPAD at $10^5$ cnt/s.

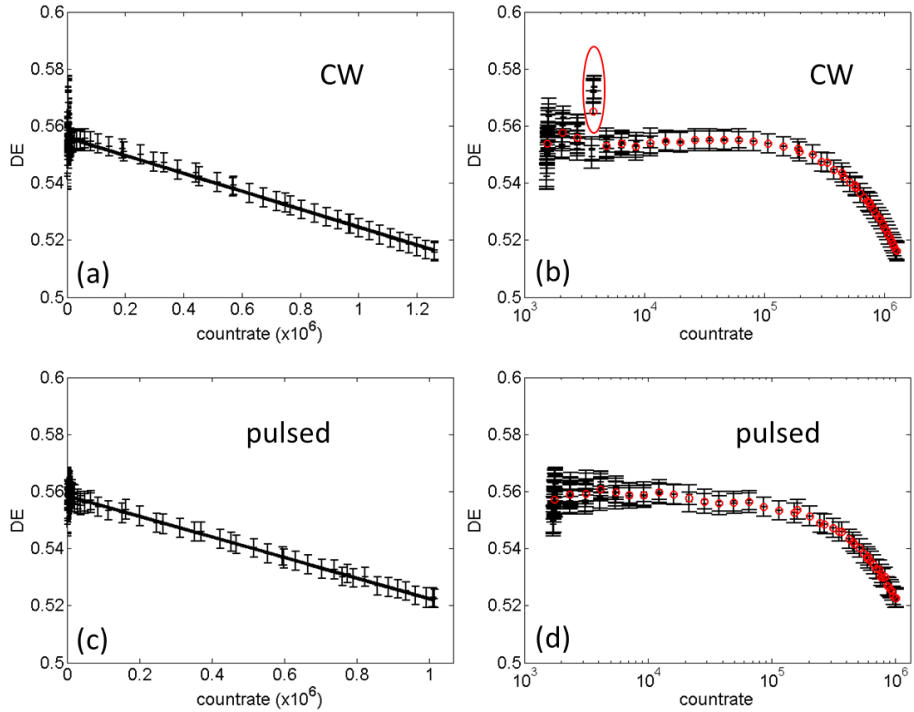

Figure 10: Measured efficiencies versus count rate for NIST8103. (a), (b) were measured with the CW laser on a linear and logarithmic count rate scale, respectively. (c), (d) were measured with the Ti:sapphire laser on a linear and logarithmic scale, respectively. The black data points and error bars represent the measured value and standard uncertainty ($k$=1) at each count rate. The black lines in (a) and (c) are linear fits to the data. The red open circles in (b) and (d) show the mean of the extracted DE versus the mean of the count rate at each VFA$_{input}$ setting.

The circled data points in Figure 10(b) show larger extracted DE values than the mean value of any surrounding data points. The reason for these outliers is a bi-stable dark count rate of that particular SPAD detector [32]. When we take several DE measurements at one attenuator setting, we only measure the dark count rate once before acquiring the actual source-photon counts for this attenuator setting. Hence, if the dark count rate changes during the measurement, the estimated DE will change. In this case, the dark count rate increased during a run and stayed high for all subsequent low count rates (<3500 cnt/s) until the SPAD was illuminated with a photon flux resulting in about $10^6$ cnt/s. This effect will manifest itself most clearly at low photon count rates. Table 4 summarizes the extracted DE results at 1 cnt/s and $10^5$ cnt/s for NIST8103 with the CW laser and Ti:sapphire laser. The mean measurement wavelength, DE at 1 cnt/s and $10^5$ cnt/s, the respective output-to-monitor ratio ($R_{out/mon}$), and lab temperature are given along with their extracted standard uncertainties.

Table 4: Results for NIST8103 at 1 cnt/s and $10^5$ cnt/s and their standard uncertainties ($k=1$), with the CW and Ti:sapphire laser

| Laser | dataset | λ (nm) | Temperature (°C) | $R_{out/mon}$ (x10⁻⁶) | DE (1 cnt/s) | DE ($10^5$ cnt/s) |
|---|---|---|---|---|---|---|
| CW | 1 | 851.73(1) | 23.30(5) | 8.116(25) | 0.5562(41) | 0.5530(42) |
|  | 2 | 851.73(2) | 23.31(6) | 8.117(25) | 0.5536(41) | 0.5507(41) |
|  | 3 | 851.73(2) | 23.31(6) | 8.129(25) | 0.5523(41) | 0.5491(41) |
|  | 4 | 851.72(1) | 23.32(5) | 8.149(25) | 0.5492(40) | 0.5464(41) |
|  | 5 | 851.72(1) | 23.31(5) | 8.170(25) | 0.5490(41) | 0.5461(41) |
| Ti: sapphire | 1 | 850.76(1) | 23.31(8) | 21.15(6) | 0.5556(40) | 0.5525(40) |
|  | 2 | 850.76(1) | 23.33(5) | 21.15(7) | 0.5587(41) | 0.5551(41) |
|  | 3 | 850.78(2) | 23.31(7) | 21.16(6) | 0.5550(40) | 0.5517(40) |
|  | 4 | 850.79(1) | 23.31(4) | 21.16(7) | 0.5574(41) | 0.5539(41) |
|  | 5 | 850.77(2) | 23.31(5) | 21.17(7) | 0.5561(40) | 0.5529(41) |

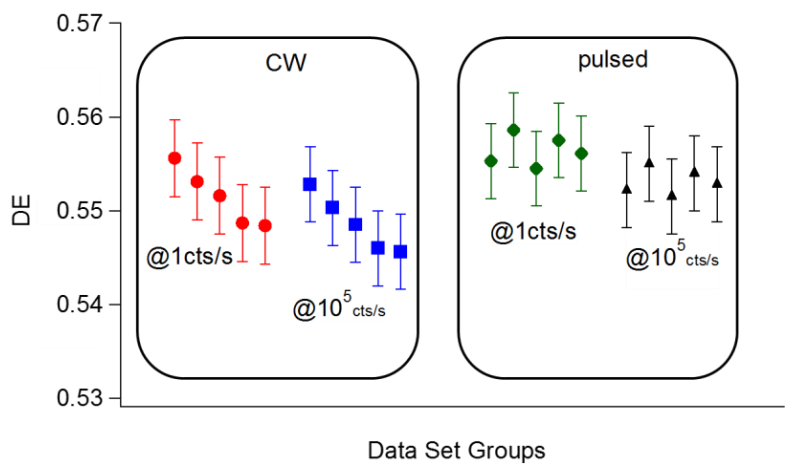

Figure 11: Data represented in Table 4. Measured DE for the NIST8103 detector with the CW laser at 1 cnt/s and $10^5$ cnt/s (solid red circles and solid blue squares, respectively) and with the pulsed Ti:sapphire laser at 1 cnt/s and $10^5$ cnt/s (solid green diamonds and solid black triangles, respectively). Error bars represent the extracted standard uncertainties ($k=1$) for each measurement.

Good setup stability and repeatability is achieved for the extracted detection efficiencies at 1 cnt/s and $10^5$ cnt/s shown in Table 4 with the pulsed Ti:sapphire laser. The data for the CW laser show a larger variation in the extracted DE for 1 cnt/s and $10^5$ cnt/s. The data show an apparent trend to lower extracted DEs for increasing dataset number. The five datasets were acquired over a 46-hour period. We speculate that the observed trend is due to drift of the *xyz*-stage during the measurements, since laser wavelength, polarization and the environmental temperature were stable. In addition, the CW laser caused fringes in the spatial response of the detector, and slight misalignments due translation stage drift will therefore have a larger impact on the extracted detection efficiencies compared to measurements with the Ti:sapphire laser. Table 6 shows the results for NIST8103 is presented in the Summary section.

## 5. Fiber-coupled single-photon detector calibration at $10^5$ cnt/s

We calibrated two fiber-coupled Si-SPAD detectors and one SNSPD at a wavelength of 851.8 nm, and one SNSPD at a wavelength of 1533.6 nm. Figure 12 shows the experimental scheme. The final free-space section of our setup is replaced with a fiber-coupled beam path to the DUT, including a fiber polarization controller (FPC) to adjust the polarization for our SNSPD measurements. When switching the fiber from PM to the DUT, we keep the DUT fiber end and PM fiber end together as close as possible to minimize issues induced by moving the fibers over large distances. For the 1533.6 nm measurements, the single-mode 851.8 nm optical fibers, FBS, and FPC are replaced with components that are single mode at 1533.6 nm.

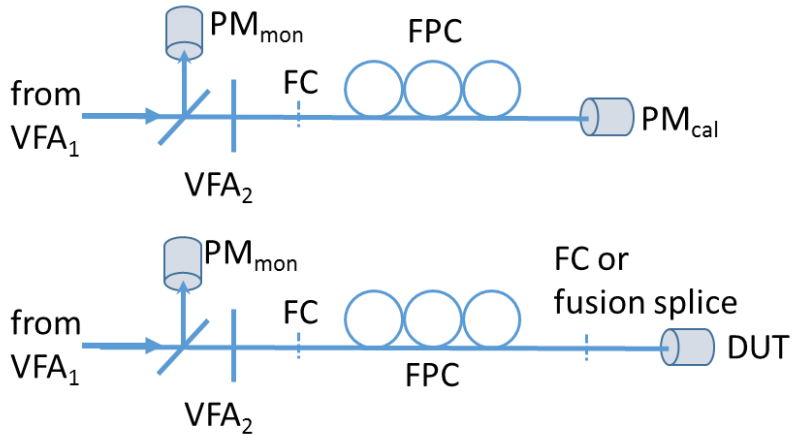

Figure 12: Fiber-coupled setup. The same beam-splitter method shown in Figure 6 is used to measure the efficiency of fiber-coupled SPDs. A fiber-coupler (FC) after the VFA connects a fiber polarization controller and the PM to measure the output-to-monitor ratio. After the output-to-monitor ratio measurement, the DUT is connected to the output of the polarization controller either through a fusion splice or fiber connector (FC).

We use a commercial optical-fiber power meter as the PM for our fiber-coupled measurements. We use a Si-based PM and an InGaAs-based PM for the two measurement wavelengths of 851.8 nm (PM(851.8 nm)) and 1533.6 nm (PM(1533.6 nm)), respectively. We calibrated each PM using the NIST service for absolute calibration and nonlinearity as described above. We use $PM_{mon}$ for both wavelengths. The FPC serves as a polarization controller for use with the SNSPDs, which generally show significant polarization sensitivity [28]. The connection to the DUT is achieved by either fusion splicing the detector fiber directly to the output of the FPC fiber, or by use of an FC/PC fiber connector. Since the FPC output fiber is uncoated, each fiber-coupled calibration requires a correction for the reflection off the output surface of the fiber leading into the PM. For both wavelengths, we use the manufacturer-specified effective single-mode index of the fiber at the given wavelength and calculated the reflection from the end of the fiber. We measure the DE as a function of count rate and perform a linear fit to the data and interpolate to extract the efficiency at $10^5$ cnt/s. The uncertainty for the DE at $10^5$ cnt/s was determined with the same method used above.

### 5.1 Fiber-coupled 851 nm calibration

Figure 13 shows the estimated detection efficiencies of three fiber-coupled SPDs versus detected count rate. Figure 13(a), (c) and (e) show our results of one run for detectors V23172, V23173 and PD9D on a linear scale, respectively, along with linear fits to the data that are used for the interpolation to $10^5$ cnt/s. Figure 13(b), (d), and (f) show the same data for all three detectors on a $\log_{10}$ scale. The red open circles show the average estimated detection efficiencies at each average count rate for a given VFA$_{input}$ setting. V23172 and V23173 are two fiber-coupled Si-SPAD detectors, similar to the free-space NIST8103 detector. However, a fixed internal lens focuses the light onto the active region of the detector such that there are no fiber-to-fiber connection misalignment issues. Thus, we directly connected the output fiber of the FPC to the two detector modules. For V23172 and V23173 we did not apply the end-reflection correction and we report the DE for photons that exit the FPC end facet. The data in Figure 13(a) shows a variation in *DE* around the solid black line. This variation is most noticeable at count rates beyond 300000 cnt/s, and the uncertainty of the mean DE predicted by the linear fit is a significant contribution to the final uncertainty of DE. We determined the influence of this variation on the result by excluding the data beyond 300000 cnt/s and performing the same analysis for extrapolation (interpolation). We found that for both extrapolation and interpolation, the difference of *DE* is within the standard uncertainty of the final result, while the uncertainty of the mean DE predicted by the linear fit is also substantially reduced.

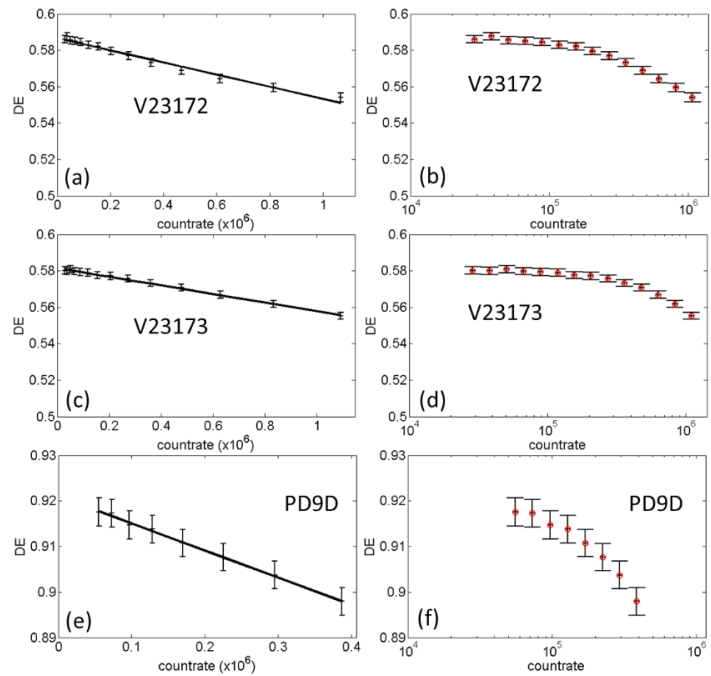

Figure 13: Estimated DE of three fiber-coupled SPDs vs. detected count rate at a wavelength of about 851.8 nm. (a) Results for detector V23172 on a linear scale. The black line represents a linear fit to the data. The black symbols and error bars represent the measured value and the standard uncertainty ($k$=1) at each count rate. (b) Results for detector V23172 on a $\log_{10}$-scale. The red open circles represent the average estimated detection efficiencies at the average count rate for each VFA$_{input}$ setting. (c) and (d) show the results for detector V23173. (e) and (f) show the results for SNSPD PD9D.

Table 5: Results for 4 detectors at $10^5$ cnt/s and standard uncertainties ($k$=1), with CW laser and fiber or free-space connection.

| Detector | Connection | dataset | λ (nm) | Temperature (°C) | $R_{\text{out/mon}}$ (x10⁻⁶) | DE |
|---|---|---|---|---|---|---|
| V23172 | direct-fiber | 1 | 851.80(1) | 22.58(5) | 34.08(10) | 0.5788(51) |
|  |  | 2 | 851.80(1) | 22.60(8) | 34.10(10) | 0.5817(47) |
|  |  | 3 | 851.80(1) | 22.63(9) | 34.12(10) | 0.5829(49) |
| V23173 | direct-fiber | 1 | 851.77(1) | 22.78(5) | 32.96(10) | 0.5792(37) |
|  |  | 2 | 851.80(1) | 22.76(5) | 34.03(10) | 0.5842(44) |
|  |  | 3 | 851.79(1) | 22.46(5) | 34.11(10) | 0.5829(39) |
| PD9D | Splice | 1 | 851.76(1) | 22.46(6) | 31.17(9) | 0.9151(50) |
|  |  | 2 | 851.76(1) | 22.66(7) | 31.17(9) | 0.9208(52) |
|  |  | 3 | 851.76(1) | 22.58(5) | 31.12(9) | 0.9177(57) |
| NS233 | Connector | 1 | 1533.62(1) | 22.84(10) | 4.984(10) | 0.8908(28) |
|  |  | 2 | 1533.62(1) | 23.09(35) | 4.978(10) | 0.8911(28) |
|  |  | 3 | 1533.62(1) | 22.86(17) | 4.984(10) | 0.8944(29) |
| NS233 | Splice | 1 | 1533.62(1) | 22.94(10) | 4.767(10) | 0.9235(30) |
|  |  | 2 | 1533.62(1) | 22.99(8) | 4.757(10) | 0.9250(30) |
|  |  | 3 | 1533.62(1) | 23.03(24) | 4.923(10) | 0.9218(29) |

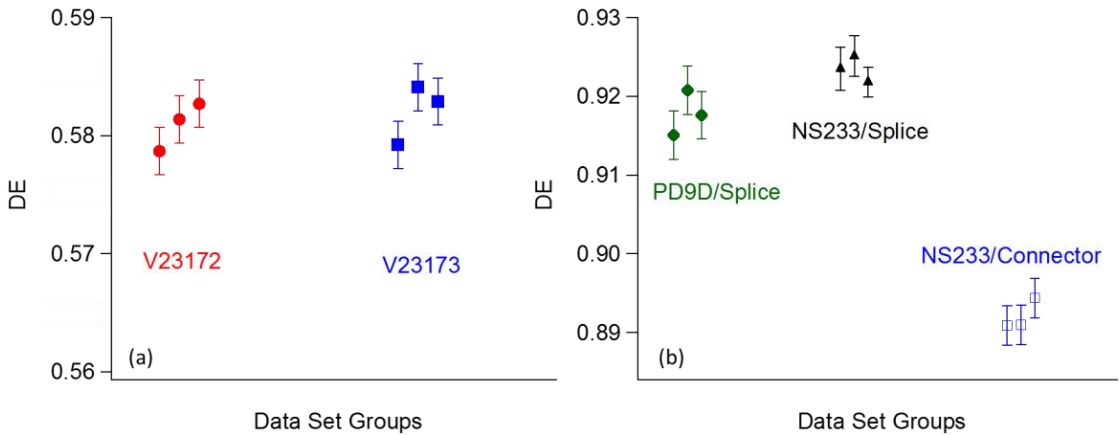

Figure 14: (a) Summary of fiber-coupled 851.8 nm measurements for detectors V23172 and V23173. (b) Summary of fiber-coupled SNSPD measurements at 851.8 nm (PD9D) and 1533.6 nm for detector NS233 using fusion splicing and FC/PC connectors to connect the DUT fiber to the output of the FBS. Error bars represent the extracted standard uncertainties ($k$=1) for each measurement.

### 5.2 Fiber-coupled 1533.6 nm calibration

Figure 15 shows the calibration results of an SNSPD (NS233) optimized for 1550 nm at a measurement wavelength of 1533.6 nm. Applying the same measurement scheme as above, we switched to a single-mode 1533.6 nm components (FBS and FPC) and fibers. The change of fibers for a different operating wavelength is straightforward and only requires switching fibers via FC/APC connectors between the output of VFA$_{input}$ and the output of the FPC.

The DE of NS233 was determined with an FC/PC fiber-to-fiber connector union (Figure 15(a) and (b)) and by fusion splicing the detector fiber to the output fiber of the FPC (Figure 15(c) and (d)). Figure 15(a) and (c) show the extracted DE versus count rate for the FC/PC connector and the fusion splice on a linear scale, respectively. The black line shows a linear fit to the data. Figure 15(b) and (d) show the data on a logarithmic scale. The red open circles represent the average of the extracted detection efficiencies at each count rate. The black data points and error bars represent the measured value and standard uncertainty at each count rate. Note the difference between the fusion-spliced and connectorized fibers is about 3.5 %, which is within the range of typical loss that we see with FC/PC connectors.

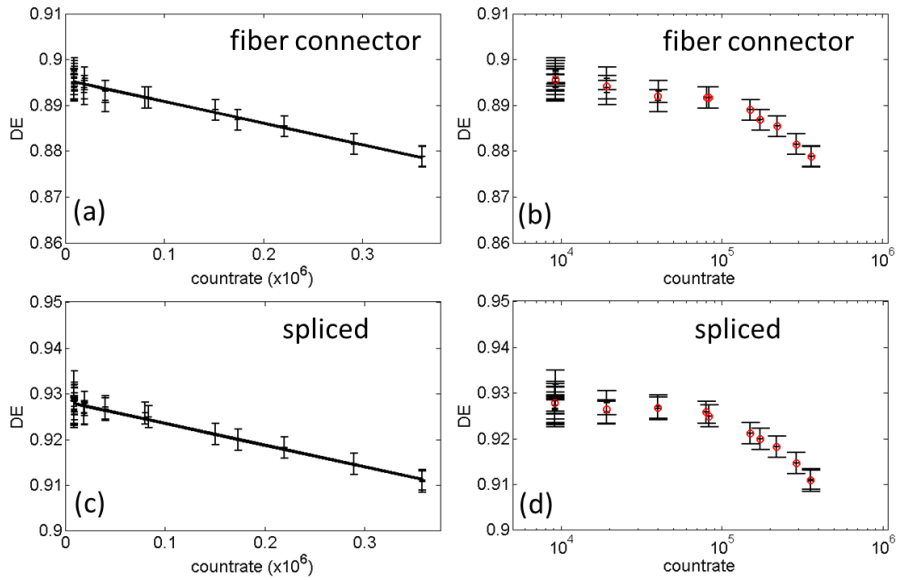

Figure 15: Estimated DE of NS233 vs. detected count rate at a wavelength of about 1533 nm. (a) Results with an FC/PC connector on a linear scale. The black line represents a linear fit to the data. The black symbols and error bars represent the measured value and standard uncertainty ($k$=1) at each count rate. (b) Results with an FC/PC connector on a log$_{10}$-scale. The red open circles represent the mean estimated detection efficiencies at each count rate. (c) and (d) show the results for detector NS233 with a fusion splice.

Table 5 and Figure 14(b) show a summary of the results obtained for both connector cases. The measured DE through a fusion splice is higher than that measured through an FC/PC connector, as we would expect. The repeatability between individual runs for both cases is comparable to the repeatability achieved for the 851.8 nm fiber-coupled measurement. Table 6 shows the results for NS233 and is presented in the Summary section.

## 6. Summary

Table 6: Summary of results for all measured SPDs and their setup configuration. Quoted are the mean measurement wavelength, number of measurements, extracted mean DEs at $10^5$ cnt/s, the 95 % coverage intervals and the relative expanded uncertainties ($k$=2).

| detector | Laser | | Fiber-coupled | | | Wavelength (nm) | Measurements | DE at $10^5$ cnt/s | 95 % coverage interval | relative expanded uncertainty ($k$=2) (%) |
| | CW Ti:sapphire | Free-space | Direct fiber connectorized | | Fusion-spliced | | | | | |
|---|---|---|---|---|---|---|---|---|---|---|
| NIST8103 | x | x | | | | 850.77 | 5 | 0.5532 | [0.5449, 0.5615] | 1.53 |
| NIST8103 | x | x | | | | 851.73 | 5 | 0.5490 | [0.5397, 0.5587] | 1.78 |
| V23172 | x | | x | | | 851.80 | 3 | 0.5811 | [0.5708, 0.5911] | 1.75 |
| V23173 | x | | x | | | 851.79 | 3 | 0.5821 | [0.5735 0.5911] | 1.56 |
| PD9D | x | | | | x | 851.76 | 3 | 0.9178 | [0.9066, 0.9292] | 1.08 |
| NS233 | x | | | x | | 1533.62 | 3 | 0.8921 | [0.8859, 0.8996] | 0.73 |
| NS233 | x | | | | x | 1533.62 | 3 | 0.9234 | [0.9171, 0.9298] | 0.70 |

Table 6 summarizes the results of this work for all detectors at a count rate of $10^5$ cnt/s. The DE and 95 % coverage interval were calculated with the NIST consensus builder [33] and linear opinion pooling for the individual measurement outcomes for each detector. Relative expanded uncertainties as low as 0.64 % are achieved in the case of a fiber-coupled SNSPD at 1533.6 nm. Whereas for the free-space measurements at 851.8 nm, the relative expanded uncertainty is 1.78 % with a CW laser. The main source of uncertainty for the free-space measurements is the uncertainty in the detector response due to laser-beam-detector alignment. For all-fiber-coupled detectors this uncertainty is not relevant but is replaced with a connector and fiber-end reflection-loss uncertainty. In this study, we were not able to compare several FC/PC connectors to establish an uncertainty associated with different commercially available fiber connectors. However, we believe that for many different of FC/PC connectors the loss uncertainty will be larger than our overall uncertainty budget. For the NS233 detector, we observe a ≈3.5 % lower system DE than when splicing the fibers. In the extreme case, an FC/PC connection may have very low losses (close to 0 %). Therefore, we speculate that this measurement already reveals a variation of at least 3.5 % in the extracted DE for the FC/PC connector method. Also, care needs to be taken when splicing fibers. Fibers of different mode field diameters will pose different losses, and an uncertainty cannot easily be estimated for a fiber combination if the loss cannot be measured beforehand. This poses a challenge when operating superconducting or other fiber-coupled detectors from which the optical fiber cannot easily be removed beforehand to measure the fiber connection/fusion loss. Table 6 also shows a difference of the expanded uncertainty when comparing the 1533.6 nm and 851 nm measurements. This discrepancy is mainly due to the nonlinearity correction applied to our measurements. The FBS method requires three nonlinearity corrections, and at 1533.6 nm the nonlinearity correction for our power meters has a standard uncertainty of at least a factor of two less than at around 851 nm.